\documentclass{optica-article}

\journal{opticajournal} 

\articletype{Research Article}
\usepackage{xcolor}

\usepackage{lineno}
\usepackage{soul}

\begin{document}

\title{Computational imaging through turbid, bubbly and flowing water near the telescope}

\author{Alexander Ingold,\authormark{1} Sabina D. Menon,\authormark{2} Aadhi Umamageswaran,\authormark{3} Alec Ikei,\authormark{2} May H. Hlaing,\authormark{2} Syed N. Qadri,\authormark{2} and Rajesh Menon\authormark{1,*}}

\address{\authormark{1} Department of Electrical and Computer Engineering, University of Utah, Salt Lake City, UT 84112, USA\\
\authormark{2} U.S. Naval Research Laboratory Remote Sensing Division, Washington DC 20375, USA.\\
\authormark{3} Department of Electrical Engineering and Computer Science, Massachusetts Institute of Technology, Cambridge, MA 02139 USA.\\}

\email{\authormark{*}rmenon@eng.utah.edu} 


\begin{abstract*}
Cameras in the surf zone or behind ship wakes look through turbid, bubbly, moving
water close to the lens, yet restoration networks are usually scored only on
displayed images. We introduce a measured dataset of a telescope viewing displayed
images and physical objects through a 25~mm water layer 45~mm in front of its
objective, under seven combinations of flow, bubbles and turbidity, with clear-water
recordings as a second reference. Because the suspension scatters light far outside
the 1.3\textdegree{} field of view, turbidity mainly removes contrast, 4.4- to
4.8-fold on a Siemens star. Restoring the mean of eight frames improves structural
similarity (SSIM) by 0.032 over a single frame, and learned fusion of the frames
adds little. One raw frame supports five-category recognition (96.8\%) as well as
restored images. Adding synthetic training pairs leaves display-image scores
unchanged but raises the correlation of a restored Siemens star with its clear-water
recording from below 0.5 to above 0.9. These networks also improve ordinary
underwater photographs without retraining, where training on measured pairs alone
fails. Scoring against the digital source or a clear-water recording reverses the
ranking of raw and restored images. These results support a simple pipeline: average
the frames, classify raw frames directly, and restore with networks trained on
measured and synthetic pairs.
\end{abstract*}

\section{Introduction}

\begin{figure}[htbp]
\centering
\includegraphics[width=\textwidth]{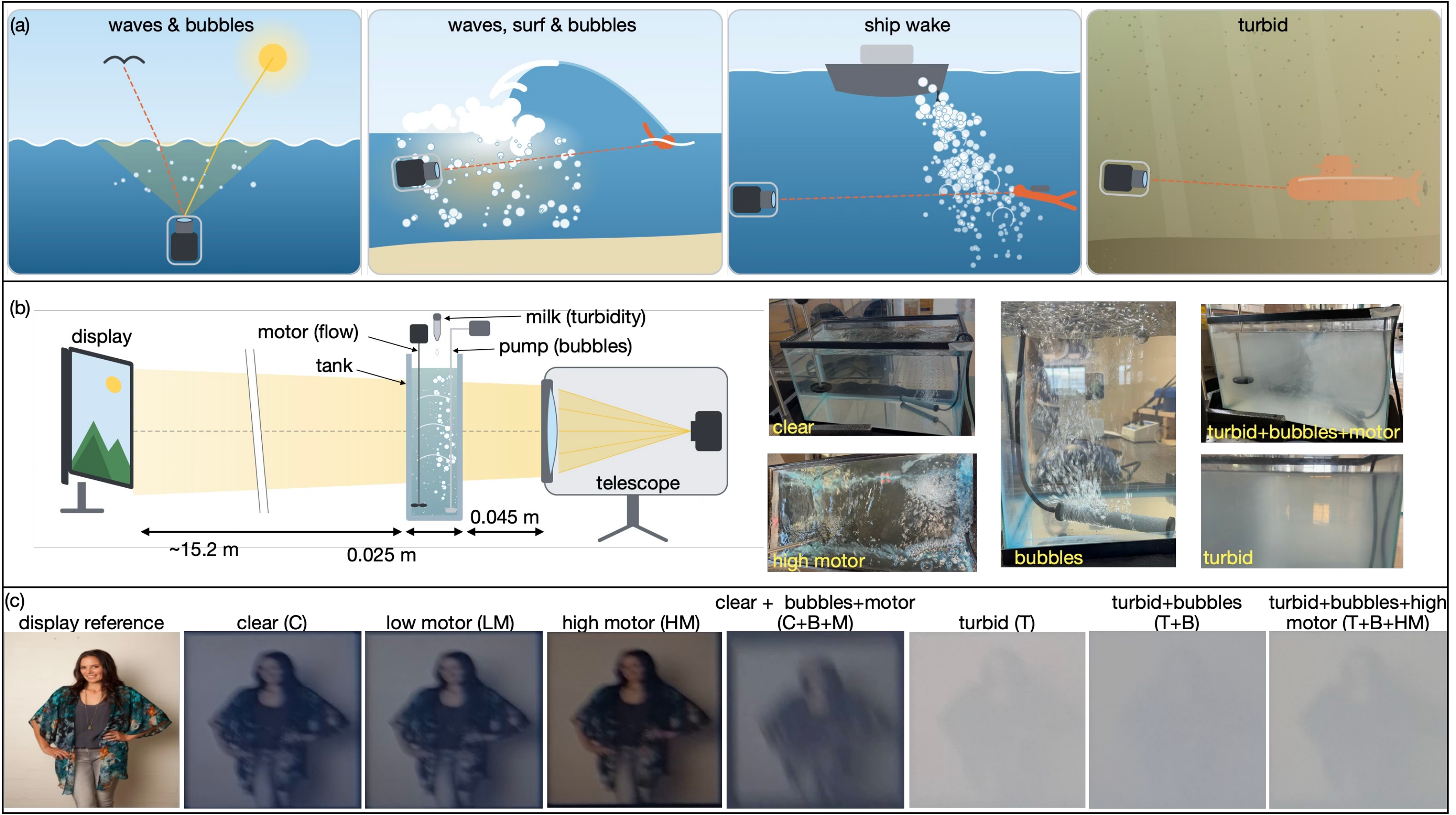}
\caption{Applications, apparatus and measurements. (a) A camera looking through
water: up through a wavy, bubbly surface, through breaking surf,
through a ship's wake, and through turbid water. (b) A Seestar S50 telescope
(50~mm aperture) views a display or physical object about 15.2~m away through a
25~mm water path whose near face is 45~mm from the objective. A motor drives flow,
an air pump adds bubbles and milk adds turbidity. Photographs show the tank in five
conditions; see also Visualizations 1--4. (c) The displayed digital image and
recorded frames of it in the seven conditions. Note that the clear image is a single frame, and degradation is due to the intrinsic telescope optics and water in the path. This image is reserved for
illustration and excluded from all quantitative results.}
\label{fig:setup}
\end{figure}

Many cameras look through water that is close to the optic and changing. A camera
in the surf zone looks through breaking waves full of bubbles and suspended
sediment. A sensor near a propeller looks through its wake. A submerged platform
looks up through a moving surface, and harbour, flood and aquaculture cameras look
through turbid water (Fig.~\ref{fig:setup}a). The output that matters may be an image, a decision about
what the scene contains, or both.

Underwater image formation is usually written as an attenuated direct image, a
forward-scattered component and a backscattered veil \cite{mcglamery1980,jaffe1990}.
Restoration methods estimate the transmission and the veil from polarization
\cite{schechner2005}, from image priors \cite{he2011dcp,drews2016udcp,peng2017blurriness},
or from a revised model in which the attenuation of the direct and backscattered
light differ \cite{akkaynak2018,akkaynak2019}. Learned enhancement is benchmarked on Underwater Image Enhancement Benchmark (UIEB), whose references are the preferred outputs of enhancement methods rather than
clear-water captures \cite{li2020uieb}; Ref.~\cite{anwar2020survey} surveys the field. These methods assume that the
medium fills the path to the scene and that it is static. A second literature treats
a moving air-water interface, whose refraction warps the image locally and
differently in each frame; restoration registers, decomposes or learns from the
sequence \cite{tian2009seeing,oreifej2011twostage,li2018turbulent,james2019nonrigid,thapa2020fluid}.
Bubbles entrained by breaking waves are well characterized as scatterers
\cite{zhang1998bubbles,terrill2001bubble,stramski2001bubbles,randolph2014bubbles},
but we found no study of imaging or restoration through bubbly, turbid water with
measured ground truth.

Learned imaging through scattering has recovered objects through diffusers, thick
slabs and dynamic media \cite{li2018speckle,lyu2019thick,sun2019dynamic}, and gating
networks adapt to a range of scattering strengths \cite{tahir2022}. Closest to our
work, Liu et al. trained a network on pairs recorded through a turbid tank with
targets shown on a display and recovered physical objects through milk and fog
\cite{liu2024realtime}. In our earlier work on fog, a network trained on real paired
images defogged real scenes \cite{pollak2024defog}, and pixel-registered fog-chamber
data combined with synthetic fine-tuning carried defogging across domains
\cite{ingold2026fog}. These studies score restoration against the
displayed or clear target. 

The optics of a thin scattering layer near the lens predict how these questions
should come out. A layer near the receiver is the worst position for blur from
small-angle scattering, because an angular deflection at the pupil maps directly to
an image displacement \cite{dror1998shower,edrei2016shower}. When the scattering
lobe is much wider than the field of view, however, the scattered light leaves the
image or spreads into a nearly uniform veil, and the medium lowers contrast by the
same factor at all fine spatial frequencies
\cite{bissonnette1992,sadot1993aerosol,dror1995aerosol}. Modulation transfer
functions of turbid media have been modelled \cite{wells1969} and measured with
point sources and sector stars \cite{hou2008,dror1998shower,berrocal2016sif}, but
not with bubbles and flow, and not in relation to learned restoration or
recognition.

If the water mainly removes contrast, three consequences follow. First, burst
methods gain over one frame when frames differ in blur, shift or local distortion
\cite{delbracio2015fba,hasinoff2016burst,wang2019edvr,bhat2021burstsr,dudhane2023burstormer,anantrasirichai2013,zhang2024tmt};
a degradation that is the same across the field and changes slowly offers little
such diversity, so the frame average should carry most of the information. Second,
recognition should not need restoration. Dehazing and enhancement do not reliably
help classification or segmentation \cite{pei2018haze,sakaridis2018,vidalmata2021},
and classifiers are more sensitive to blur and noise than to contrast loss
\cite{dodge2016quality}, although corruptions absent from training still lower
accuracy \cite{hendrycks2019corruptions}; we therefore train the classifier on the
degraded frames. Third, because the measurement already holds the structure, a
restoration network can add tone and suppress noise, or it can invent structure
that standard metrics do not penalize \cite{bhadra2021hallucination}. Which of the
two happens depends on the training data.

We test these predictions on water close to the optic. Seven combinations of flow,
bubbles and turbidity are recorded through a tank 45~mm in front of a telescope
objective (Fig.~\ref{fig:setup}b), with displayed images for paired evaluation and
physical objects for transfer. Our contributions are:
\begin{enumerate}
\item A dataset of water close to the lens under flow, bubbles and turbidity,
recorded with two kinds of reference: the displayed digital image, and clear-water
recordings of the same displayed images, and of physical objects.
\item A Siemens-star measurement showing that turbidity cuts contrast four- to
five-fold with lesser image blur, with and without bubbles and flow, as expected
for a scattering lobe much wider than the field-of-view.
\item A test of learned burst fusion against a network trained on the frame mean,
across three architectures, showing that eight frames add little beyond their
average.
\item Synthetic training pairs, extending our defogging approach
\cite{ingold2026fog} to dynamic water, that let a network trained on displayed
images transfer to physical objects without inventing structure. At a fixed
architecture and training budget they leave scores on displayed images unchanged,
but raise the correlation of a restored Siemens star with its clear-water recording
from below 0.5 to above 0.9.
\item Category recognition from one raw frame (96.8\%) that matches recognition from
restored images, and a demonstration that the choice of reference (digital source
or clear-water recording) reverses the ranking of raw and restored images.
\end{enumerate}

\section{Methods}

\subsection{Apparatus and data}

Static images shown on a laptop display, or physical objects placed at the display
position, are imaged by a Seestar S50 telescope (ZWO) with an objective of aperture
diameter $D = 50$~mm and focal length $f = 250$~mm onto a Sony IMX462 color sensor
($1920\times1080$ pixels, 2.9~\textmu m pitch). The object
distance, measured with a laser range finder, is 15.2~m; readings in different
sessions ranged from 15.2 to 15.6~m. A glass tank with a 25~mm water path sits with
its near face 45~mm from the objective.

The seven conditions are clear (C), low motor (LM), high motor (HM), clear water
with bubbles and motor (C+B+M), turbid (T), turbid with bubbles (T+B), and turbid
with bubbles and high motor (T+B+HM). A submerged motor at two speeds drives the
flow, an air pump adds bubbles, and milk in water sets the turbidity (Section~S8.2). We do not assign calibrated optical depths; the contrast of a Siemens
star (Section~\ref{sec:contrast}) serves as an operational measure of severity.

Each displayed image is assigned to one of the
training, tuning or evaluation sets and keeps that role in every condition, so no
evaluation image is seen in training under any condition. Paired restoration uses
2,648 training, 365 tuning and 329 evaluation recordings of 580, 76 and 72 images.
Recognition uses five categories (apparel, artwork, cars, dishes and furniture), with 2,758 training, 372 tuning and 345 evaluation recordings under the same identity roles; the additional recordings have verified category labels but no admitted paired restoration target. From each recording, eight frames are cropped to the displayed image and resampled by area
averaging to $256\times256$ RGB pixels (Section~S8.2). Intensities are the camera's encoded 8-bit
values divided by 255, without radiometric linearization.

Twenty-four physical-object recordings are used only for evaluation. Six objects (a
Siemens star, a bar target, a dot grid, a chain, an emblem and a lettered plaque)
are recorded in T and T+B+HM conditions. Four objects (the Siemens star, the dot grid, and the
bar targets are recorded in C, C+B, and C+B+HM conditions. C+B
and C+B+HM are object-only conditions. The clear-water and turbid object recordings
come from different sessions, so their image scales differ by up to 8\%; Section~3.3
describes the registration.

\subsection{Image-formation model}

Let $O_c(\mathbf{x})$ be the scene in color channel $c$ at image position
$\mathbf{x}$, and $I_c(\mathbf{x},t)$ the recorded frame at time $t$. We write
\begin{equation}
I_c(\mathbf{x},t) = \alpha_c(t)\,\bigl[h_t \ast O_c\bigr](\mathbf{x}-\mathbf{s}_t)
 + V_c(t) + n_c(\mathbf{x},t),
\label{eq:model}
\end{equation}
where $\alpha_c$ is the transmitted fraction, $h_t$ the point-spread function,
$\ast$ convolution, $\mathbf{s}_t$ a global image shift, $V_c$ a veil that is
uniform across the field, and $n_c$ noise.

Two properties of the geometry justify this form. First, the water mid-plane lies
57.5~mm from the objective, under 0.4\% of the object distance, so the beams from
different field points overlap almost entirely in the water: at the mid-plane, the
beams of diameter $D$ from the center and the edge of the field share all but 1.6\%
of their footprint. The degradation is therefore the same across the field, and
$h_t$, $\mathbf{s}_t$ and $V_c$ do not depend on $\mathbf{x}$.

Second, the position does not by itself prevent blur; a layer near the receiver is
the least favorable position for small-angle scattering \cite{dror1998shower}. Blur
is reduced (but not eliminated) because the scattering lobe is wide compared with the field. Light
scattered in the water at up to 0.30~rad still enters the objective, but only light
traveling within $\pm 8.3$~mrad of the axis ($\pm 11.1$~mrad in air) reaches the
sensor. For a Henyey-Greenstein phase function with anisotropy $g = 0.9$, typical of
milk in water~\cite{narasimhan2006dilution}, only 0.33\% of singly scattered light
is deflected by less than this angle, and the fraction stays below 1.3\% for any
$g \le 0.95$. Scattering therefore removes light from the image, or spreads it over
angles much larger than the field, rather than forming a halo. Light from the
surroundings, scattered by the water toward
the objective, adds a nearly uniform veil. Turbidity should therefore lower
$\alpha_c$ and raise $V_c$ while leaving $h_t$ close to its clear-water form, which
is the plateau regime of the aerosol modulation transfer function
\cite{bissonnette1992,sadot1993aerosol}.

Flow and bubbles act differently. Bubbles near the pupil are millimeter-scale
refracting and occluding elements that move through the beam, and flow tilts and
distorts the water; both change $\mathbf{s}_t$ and $h_t$, identically across the
field, without adding a veil. A burst method can still gain from such changes if
$h_t$ varies strongly from frame to frame, by favoring at each spatial frequency the
frames that pass it best \cite{delbracio2015fba}. The gain should be small when
$h_t$ varies little over the eight frames, which Section~\ref{sec:burst} tests.
Supplement~1, Section~S8.1 gives the derivation.

\subsection{Restoration, baselines and recognition}

The primary restorer is a joint-RGB Restormer \cite{zamir2022restormer}. It is
trained for 65,536 updates to map recorded frames to the displayed digital image,
with a loss that combines Charbonnier pixel error, SSIM and image-gradient error
(Section~S8.3). One update in four uses a synthetic pair instead of a
measured one, following our defogging work \cite{ingold2026fog}. Each synthetic
pair takes a clean $256\times256$ natural image crop from the frozen COCO test2017 source roster as the target and
degrades it eight times according to Eq.~(\ref{eq:model}): a channel-dependent
transmission $\alpha_c$, an additive veil $V_c$, a smooth gain that varies slowly
across the image, Gaussian blur for $h_t$, and weak independent noise, with the
parameters drifting by 2\% across the eight views. Degradation strength ranges from
none to strong, and the parameter ranges are fitted to training recordings only.
The generator does not model bubbles or refraction explicitly, and its parameters
are not calibrated optical depths.

Separately trained models receive either the eight frames (burst) or their
arithmetic mean repeated into all eight inputs (trained mean), so only the input
differs. EDVR \cite{wang2019edvr} and NAFNet
\cite{chen2022nafnet}, each with single, mean and burst inputs, are architecture controls. A Restormer
trained on measured pairs only, with the same budget, isolates the synthetic data.
The primary restoration comparisons use seed 1337, while Section~S4 includes the cuurent-dataset two-seed Restormer comparison at completed Seeds 1337 and 4099; checkpoints are selected by the highest tuning-set SSIM\cite{wang2004ssim}.

The baseline with no training is a per-channel contrast stretch of the eight-frame
mean: subtract the 1st percentile, divide by the range to the 99th percentile and
clip to $[0,1]$. When $h_t$ is close to a delta function and the scene spans dark
to bright, this undoes $\alpha_c$ and $V_c$ in Eq.~(\ref{eq:model}) up to a scale.
It serves as a test of what the measurement contains rather than as a restoration
method.

Recognition uses an ImageNet-pretrained ResNet-50 with a five-category output, trained for 1,000 updates on pooled training data from all seven conditions at seeds 1337, 2027 and 4099. Its inputs are one raw frame, the eight raw frames
(frame embeddings averaged), or the output of a restorer;
Section~S3 and Table~S4 lists all 10 inputs. Class probabilities are averaged over the
recordings of each image within a condition, which gives per-condition accuracy,
and then across conditions, which gives one label per image.

\begin{figure}[htbp]
\centering
\includegraphics[width=1\textwidth]{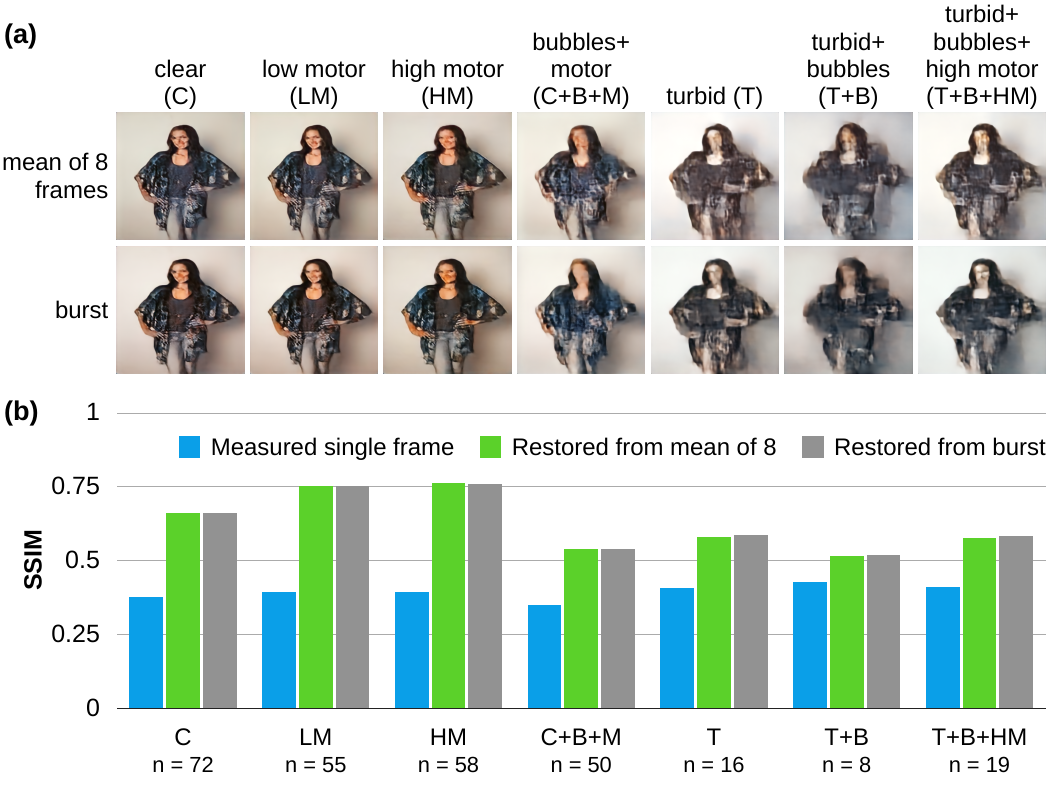}
\caption{Restoration of display images. (a) The reserved image restored by networks trained on
the frame mean and on eight frames, in each condition. (b) SSIM in reference to the displayed
image: a measured single frame, restored from mean of 8 frames, and restored from burst of frames (video); $n$ is the number of evaluation images. See also Visualizations 5--11.}
\label{fig:display}
\end{figure}

\section{Results}

\subsection{Eight frames add little beyond their average}
\label{sec:burst}

We score restoration by SSIM against the displayed digital image, computed with a
$7\times7$ uniform window on RGB values in $[0,1]$. Peak signal-to-noise ratio
(PSNR) and mean absolute error (MAE) are given in Table~S2. Scores average the
recordings of each image within a condition, then weight images equally. Paired
95\% intervals come from 10,000 bootstrap resamples of evaluation images, stratified
by category.

Across the seven conditions, the measured frame scores 0.3848 SSIM and the plain
average of eight frames 0.3878 (Table~S2). A Restormer trained on that average
reaches 0.6756; one trained on the eight frames reaches 0.6753, a burst-minus-mean
difference of $-0.0003$ (interval $[-0.0028, 0.0018]$). Training accounts for 99\%
of the gain from the measured frame to the burst network. EDVR shows the same
pattern, with trained averaging ahead by 0.0099 (interval $[0.0074, 0.0126]$). A
Restormer trained on single frames reaches 0.643, so averaging eight frames
before restoration adds 0.0322 SSIM (95\% paired interval [0.0283, 0.0361]). Learned burst fusion provides no clear further gain for Restormer.

Fidelity varies more with condition than with input (Fig.~\ref{fig:display}b).
Flow in clear water costs little: LM and HM score 0.750 and 0.762, above C at 0.661,
which pools two clear acquisitions. The trained mean score is 0.540 in C+B+M and lowest in T+B, 0.516; the three turbid conditions fall between 0.516 and 0.586. In every
condition the burst and trained-mean networks are within 0.008 of each other.

\subsection{Restoring images of real objects}
\label{sec:contrast}

\begin{figure}[htbp]
\centering
\includegraphics[width=\textwidth]{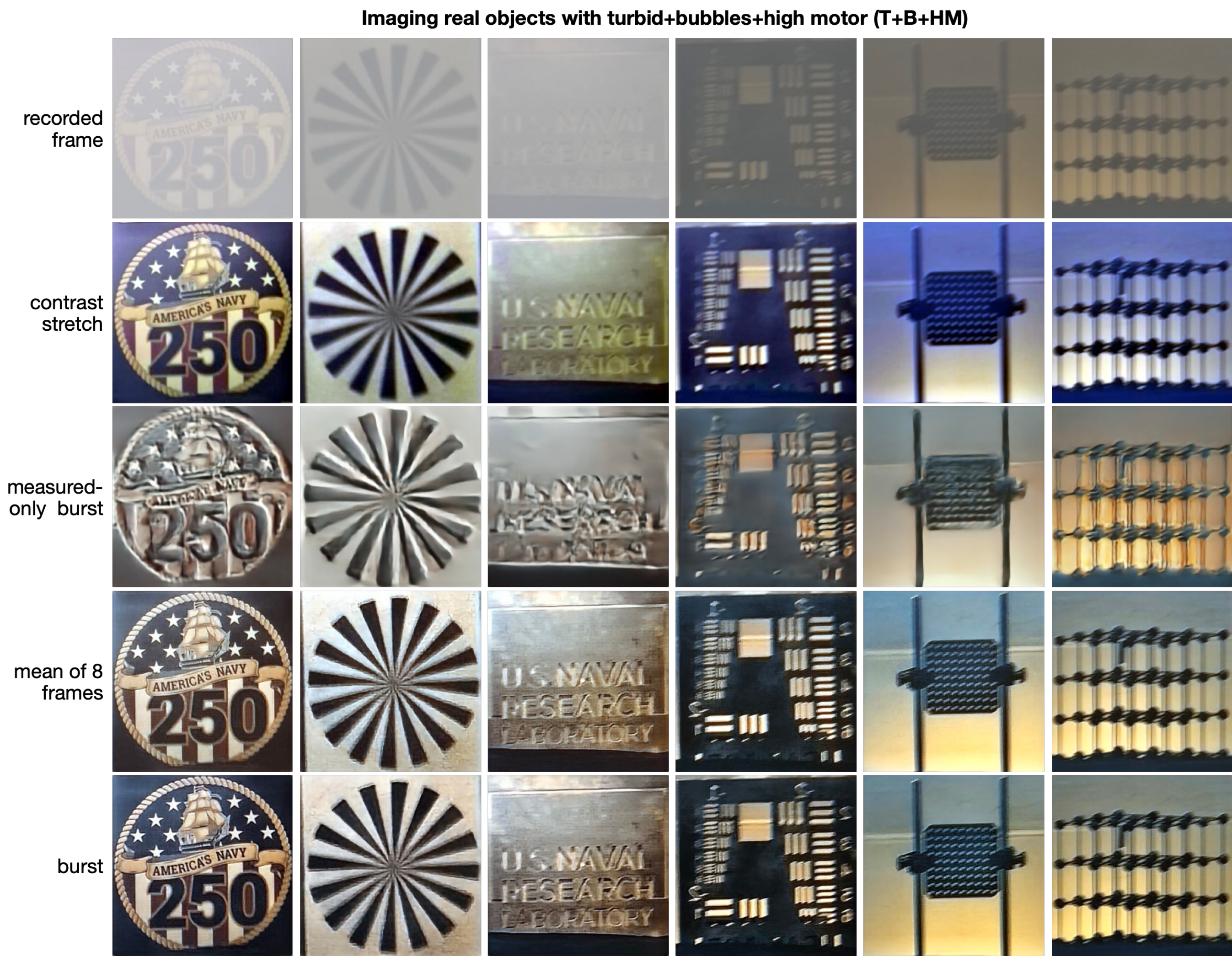}
\caption{Physical objects through turbid water with bubbles
and high motor (T+B+HM). Rows: recorded frame; contrast
stretch of the eight-frame mean, no training; Restormer trained on measured data
only; Restormer trained with synthetic data on the frame mean; the same on all
eight frames. See Visualizations 12 and 13.}
\label{fig:objectsTBHM}
\end{figure}

Physical objects recorded through turbid water appear as a near-uniform gray
(Fig.~S6), and bubbles and flow degrade them further
(Fig.~\ref{fig:objectsTBHM}). A contrast stretch of the eight-frame mean, with no
training, reveals the lettering, the bars of the resolution target and the spokes
of the Siemens star. The trained networks give cleaner color and less noise.

A Siemens star makes this quantitative (Fig.~S5;
Section~S7.1). We measure the spoke modulation $m$, the amplitude of the
azimuthal Fourier component at the spoke frequency divided by the mean intensity,
as a function of radius and hence, of spatial frequency. With the center estimated from the recorded eight-frame
mean, outer modulation falls from 0.407 in C to 0.092 in T, a 4.4-fold
loss, and from 0.277 in C+B+HM to 0.058 in T+B+HM, a 4.8-fold loss. All recordings
keep the 16 spoke cycles. The
resolution through turbid water is set by the optics and the noise, as
Eq.~(\ref{eq:model}) predicts. Section~S7.1 gives the detailed analysis and sensitivity checks.

The stretch tests what the measurement contains; it is not a substitute for
restoration when fidelity matters. On the reserved image of Fig.~\ref{fig:setup}c,
the stretch raises SSIM in C, LM and HM (from 0.46--0.51 to 0.57--0.58) but lowers
it in T, T+B and T+B+HM (from 0.55--0.56 to 0.21--0.25), because it amplifies the
noise by the same factor as the signal and does not reproduce the display's tone.
The burst network reaches 0.61 to 0.66 in the turbid conditions.

\subsection{Synthetic training data lets restoration transfer to physical objects}
\label{sec:synthetic}

Physical objects have no digital target, so we compare each output with the
clear-water recording of the same object. Each image is registered to the
clear-water eight-frame mean by a search over scale (0.92 to 1.08), which absorbs
the change in magnification between sessions, followed by sub-pixel phase
correlation. We report the Pearson correlation of luminance over the overlap.
Correlation ignores the gain and offset ($\alpha_c$, $V_c$) of Eq.~(\ref{eq:model}),
so it measures agreement in structure rather than in tone. See Section~7.2 for details.

Adding synthetic pairs to training, as in our defogging work \cite{ingold2026fog},
is what lets restoration transfer from displayed images to physical objects. On
displayed images the effect does not show: the measured-only and mixed Restormers
score 0.6759 and 0.6753 SSIM, a mixed-minus-measured difference of $-0.0006$
(interval $[-0.0033, 0.0020]$). On physical objects it is large
(Table~S7). Through turbid water, the recorded eight-frame mean correlates
with the clear-water recording at 0.92 to 0.94 for the Siemens star, 0.73 to 0.81
for the bar target and 0.52 to 0.64 for the dot grid. With synthetic pairs, the
networks stay within 0.05 of these values while giving cleaner tone and less noise.
Without them, the network falls to 0.35 to 0.47 on the star and 0.52 to 0.62 on the
bar target, and it fills uniform regions with grainy texture and distorts lettering
(Fig.~\ref{fig:objectsTBHM}). The same pattern holds in
clear water with bubbles, where the recorded mean correlates at 0.92 to 0.99 and the
measured-only network at 0.48 to 0.88. No network agrees with the clear-water
recording more than 0.03 better than the recorded mean does: restoration improves
tone and noise with similar structural agreement in these physical-object examples. A score on
displayed images therefore does not predict behavior on physical objects. A likely
reason is that measured pairs teach the network two things, to undo the water and
to map the camera's view of a display back to the digital file, and the second does
not apply to real surfaces. Synthetic pairs contain no display or camera response,
so they anchor the network to undoing Eq.~(\ref{eq:model}). 

\subsection{Transfer to ordinary underwater photographs}
\label{sec:uieb-transfer}

\begin{figure}[htbp]
\centering
\includegraphics[width=\linewidth]{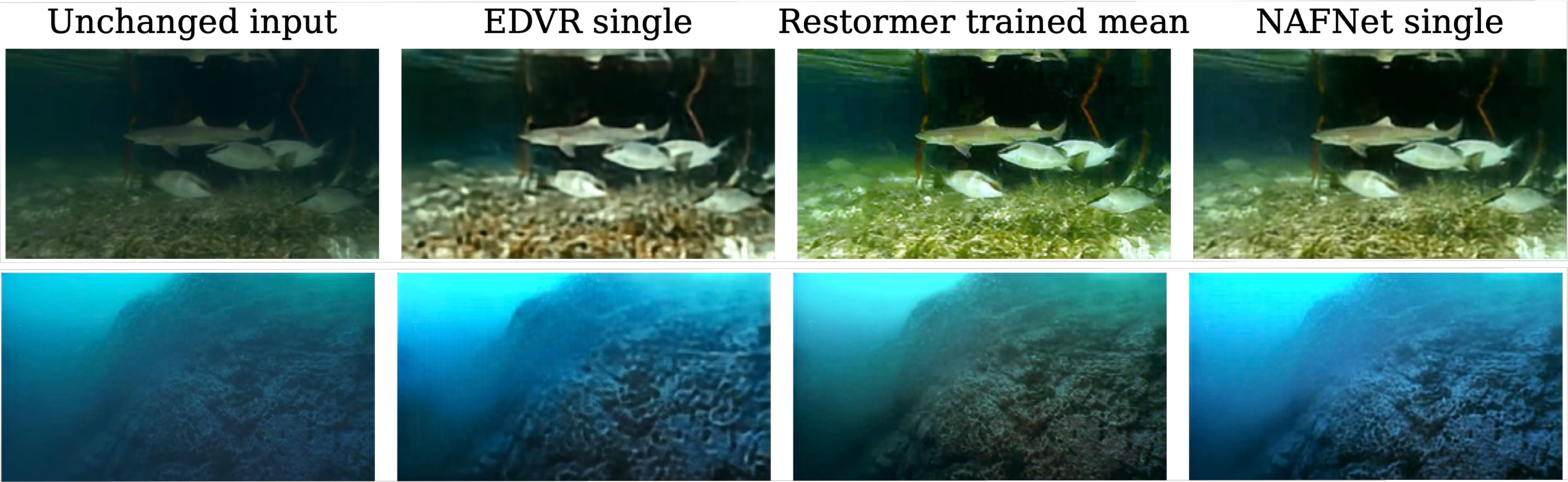}
\caption{Restoration of ordinary underwater photographs without retraining. Two
difficult UIEB photographs that have no reference images: (a) a dark fish scene and
(b) a reef slope. Columns: unchanged input, and outputs of EDVR (single-frame
input), Restormer (trained mean) and NAFNet (single-frame input), all trained with
synthetic pairs. For each architecture, the model shown is the one with the highest
mean SSIM on the 890 referenced UIEB images. Each photograph is repeated as all
eight input frames. All panels use the same display range, without brightness or
color adjustment.}
\label{fig:uieb}
\end{figure}

As a further test of transfer, we applied the restorers, without retraining, to
photographs from the Underwater Image Enhancement Benchmark (UIEB)
\cite{li2020uieb}. These are ordinary underwater scenes in which water fills the
path to the scene, far from the geometry of our tank. Each photograph is resized to
at most 256 pixels on its long edge and repeated as all eight input frames, so the
test probes single-image transfer rather than temporal fusion. The networks trained
with synthetic pairs make fish and reef structure more visible in difficult
photographs (Fig.~\ref{fig:uieb}). On the 890 UIEB images that have
references, they raise SSIM from 0.760 for the unchanged input to between 0.765 and
0.846, with the trained-mean Restormer highest (Supplement~1, Table~S3). The
Restormer trained on measured pairs only falls to 0.359, well below the unchanged
input, which repeats on a new domain the failure seen on physical objects
(Section~\ref{sec:synthetic}). The per-channel contrast stretch scores higher than
any network (0.874). UIEB references are enhanced images selected by human raters
rather than clear-water captures, so these scores measure agreement with a
preferred enhancement rather than fidelity; the ranking is nonetheless consistent
with a degradation that is mainly a loss of contrast.

\subsection{Recognition works on raw frames}

A classifier reading one raw frame identifies the category of 96.76\% of evaluation
images (standard deviation 1.60 over three classifier seeds). Eight raw frames give
96.76\%, and the burst-restored image 97.22\% (Fig.~S3). Across all 10
inputs, accuracy ranges from 95.4\% to 99.5\%, a spread of about three images out
of 72. 
Relative to raw single-frame recognition, raw eight-frame recognition changes accuracy by 
0 percentage points (95\% paired identity-bootstrap interval -1.85 to 1.85), and burst-restored recognition changes it by 
0.46 points (-2.31 to 3.24). These intervals resample identities and condition on the three fitted classifier seeds.

The one condition where restoration helps is clear water with bubbles and motor
(C+B+M). Accuracy rises from 80.7\% for raw single frames to 91.3\% for burst
restoration, a paired difference of $+10.7$ percentage points (95\% interval $+3.3$
to $+18.7$; $+0.7$ to $+22.0$ after Bonferroni adjustment for seven conditions) over
50 images in five categories. The interval resamples images, not training seeds.
C+B+M is also a condition with low restoration fidelity: in clear water,
bubbles change the point-spread function rather than adding a veil. Conditions
differ in the number of images and categories present; T+B has 9 images in 2
categories and is not interpreted.

\section{Discussion}

The geometry and the scatterer account for the results. We approximate the degradation as spatially uniform because beams from different field points overlap almost entirely in the near-lens water. Because
the turbid suspension scatters at angles much wider than the field, turbidity
removes light from the image, light scattered in from the surroundings adds a
uniform veil, as in the plateau of the aerosol modulation
transfer function \cite{bissonnette1992,sadot1993aerosol}. Larger scatterers, with a
narrower lobe, or a wider field of view would move the system toward the
blur-dominated regime of the shower-curtain effect \cite{dror1998shower}. Within this approximation, bubbles
and flow move and reshape a shared point-spread function. The small gain from burst fusion suggests limited useful diversity 
over eight frames. Multi-frame methods gain when the blur changes strongly from frame
to frame, as with camera shake \cite{delbracio2015fba}, or varies across the field,
as with a wavy surface or atmospheric turbulence
\cite{oreifej2011twostage,anantrasirichai2013,zhang2024tmt}. The same picture
explains why a contrast stretch reveals physical objects and why a classifier
trained on raw frames reads them as well as restored images, in line with findings
that dehazing does not reliably aid recognition \cite{pei2018haze,vidalmata2021}.

This also defines what restoration is for in this regime. A trained network improves
tone and suppresses the noise that a stretch amplifies, and it raises recognition
when bubbles change the point-spread function in clear water. Synthetic training pairs reduce
invented structure on physical objects, a failure that scores on displayed images
do not reveal \cite{bhadra2021hallucination}. Together with our results for fog
\cite{ingold2026fog}, this suggests a general recipe for learned restoration through
scattering media: train on measured pairs for realism, and add synthetic pairs built
from a simple attenuation-and-veil model so the network learns to invert the medium
rather than the display.

How restoration is scored matters as much. For the 279 evaluation recordings of 70
images that also have a separate clear-water recording (the unrestored eight-frame average), we score each output against that recording as well as against the
digital source. Against the digital source, the burst network beats the eight-frame
mean by 0.294 SSIM and 11.25~dB PSNR. Against the clear-water recording the ranking
reverses: the unrestored 8-frame average scores 0.842 SSIM and 27.24~dB, the burst network 0.449 and
8.83~dB, a difference of $-18.4$~dB (interval $[-19.9, -16.9]$). A network trained
against the digital source learns to undo the tone and blur of the display and
camera; a clear-water recording keeps them. Tone error also hides the benefit of
averaging: eight frames improve PSNR over one frame by 0.37~dB (interval
$[0.18, 0.59]$) against the clear-water recording, but by $-0.005$~dB (interval
$[-0.018, 0.007]$) against the digital source. Neither reference alone measures
fidelity. A study that trains and scores only against displayed images
\cite{liu2024realtime} measures part of the display and camera response along with
the water; reporting both references, or a score that ignores tone, as we use for
physical objects, separates the two.

For the tested physical objects, averaging the frames and stretching contrast reveal substantial structure. Classifying directly is useful when a label is the goal,
adding a network trained with synthetic data when a clean image is needed. The near-lens geometry occurs in: a bubbly or
wavy surface layer above an upward-looking camera, a wake or bubble curtain at a
camera's port, or a turbid layer at a window (Fig.~\ref{fig:setup}a). When
scattering is spread along a long path, or the lobe is narrower than the field,
forward-scattered light reaches the sensor and Eq.~(\ref{eq:model}) no longer holds;
that regime is the subject of follow-on work. The targets are static; one turbidity
level and one bubble source are tested; the physical-object results rest on 24
recordings of six objects, with turbid and clear-water recordings from different
sessions; the evaluation images were inspected during earlier analyses.

\section{Conclusion}

Moving, bubbly and turbid water near a lens mainly hides contrast; much of the
structure remains in the measurement. A contrast stretch reveals it in physical
objects whose recorded contrast has fallen more than fourfold. Restoring the mean of
eight frames captures the multi-frame gain, and learned fusion adds little.
Synthetic training pairs let networks trained on displayed images transfer to
physical objects and to ordinary underwater photographs, with less invented
structure than training on measured pairs alone. Recognition works on raw frames,
and the choice of reference decides whether restoration appears to help.

These results suggest how to build and test cameras that look through dynamic water.
Frame averaging and contrast stretching cost little, and because a classifier reads
raw frames, buoys, drones and underwater vehicles could recognize objects through
surf, wakes and turbid water without running restoration on board. Because the
degradation stays contrast-limited only while the scattering lobe is wider than the
field, a narrow-field optic behind the water keeps the processing simple. Finally,
measured pairs combined with synthetic pairs from a simple attenuation-and-veil
model, evaluated against references that separate tone from structure, offer a
recipe for learned imaging through other dynamic scattering media, such as rain on a
window or fog on a windshield, where measured ground truth is scarce.

\begin{backmatter}
\bmsection{Funding} A.I. and R.M. acknowledge funding from US Department of Energy grant \#55801063. S.D.M. acknowledges support through the Science and Engineering Apprenticeship Program (SEAP), administered by the American Society for Engineering Education (ASEE) on behalf of the Office of Naval Research, Department of the Navy. 

\bmsection{Acknowledgment} The support and resources from the Center for High Performance Computing at the University of Utah are gratefully acknowledged.

\bmsection{Disclosures} The authors declare no conflicts of interest. The authors used LLM solely for grammar refinement and organizational suggestions. 

\bmsection{Data Availability} Data and code underlying the results are available in the Supplementary Document and will also be available on the Github page after publication.

\bmsection{Supplemental document}
Supplement 1 provides acquisition and frame-sampling details, restoration and recognition protocols and results, the current dataset two-seed Restormer comparison, independent clear-reference and temporal analyses, physical-object and Siemens-star comparisons, and supporting image-formation and synthetic-training methods.

\end{backmatter}

\bibliography{sample}

\end{document}